\documentstyle[12pt]{article} 

\def\lsim{\mathrel{\lower2.5pt\vbox{\lineskip=0pt\baselineskip=0pt 
           \hbox{$<$}\hbox{$\sim$}}}} 
\def\gsim{\mathrel{\lower2.5pt\vbox{\lineskip=0pt\baselineskip=0pt 
           \hbox{$>$}\hbox{$\sim$}}}}

\begin{document}

\begin{flushright}
hep-ph/9804329 \\
KGKU-98-01 \\
April 1998 
\end{flushright}

\vspace{15mm}

\begin{center}
{\large  \bf 
The CKM Matrix and Its Origin }
\end{center}

\vskip 15mm

\begin{center} 
Takeo MATSUOKA 

\vspace{10mm}

{\it 
Kogakkan University, Nabari \\
           Mie, 518-0498 JAPAN 
}
\end{center}

\vspace{35mm}

\begin{abstract}
In the context of the supersymmetric unification model in 
which the massless sector contains extra particles beyond 
those in the minimal supersymmetric standard model, 
we obtain mixings between quarks (leptons) and the extra 
particles which are closely intertwined with Yukawa hierarchies. 
With the assumption that the unification gauge group $G$ 
includes $SU(2)_R$, 
it is shown that the non-trivial texture of the CKM matrix 
originates from the extra-particle mixings. 
The CKM matrix of quarks emerges as a consequence of the mixings 
between the down-type quarks and colored Higgses, 
both of which are $SU(2)_L$-singlets. 
On the other hand, the CKM matrix of leptons is due to the mixings 
stemming from the seesaw mechanism with the hierarchical Majorana 
mass matrix of right-handed neutrinos. 
\end{abstract}

\newpage 
\section{Introduction}

The characteristic structure of the fermion spectra and flavor mixings 
likely has an important connection with the gauge symmetry 
and matter content above the gauge unification scale $M_U$. 
Unless the unification gauge group $G$ contains $SU(2)_R$, 
the textures of interaction terms for up-sector and down-sector 
of quarks (leptons) must be distinct from each other. 
Then, the diagonalization matrices of fermion mass matrices 
necessarily distinguish the up-sector from the down-sector. 
In this case the non-trivial structure of 
the Cabbibo-Kobayashi-Maskawa (CKM) matrix can essentially 
be traced back to the underlying theory above the scale $M_U$ 
such as the string theory. 
Alternatively, if the gauge group $G$ is large enough to 
contain $SU(2)_R$ as well as the standard model gauge group 
$G_{SM}=SU(3)_c \times SU(2)_L \times U(1)_Y$, 
the up-sector of quarks (leptons) shares interaction terms 
with the down-sector. 
Namely, above the scale $M_U$ the disparity between the 
diagonalization matrices for the up-sector and the 
down-sector does not emerge. 
This disparity occurs only below the scale $M_U$ 
purely within the effective field theory.  
In a wide class of supersymmetric unification models, 
the massless sector contains extra particles beyond those 
in the minimal supersymmetric extension of the standard 
model (MSSM), and then there may occur extra-particle mixings 
such as between quarks (leptons) and colored Higgs 
fields (doublet Higgs fields). 
As pointed out in a previous paper,
\cite{QCKM} 
it is possible that the extra-particle mixings significantly 
affect fermion masses and flavor mixings. 
Therefore, in order to study fermion masses and flavor mixings, 
it is important to specify what kinds of extra particles appear in 
the effective field theory and to study what kinds of 
extra-particle mixings take place. 
In this paper we consider an $E_6$-type model, 
which is inspired by the level-one string theory.

The hierarchical structure of fermion masses strongly suggests 
the existence of some kinds of the flavor symmetry in the 
effective field theory above the scale $M_U$. 
If this is the case, 
Yukawa couplings arise from non-renormalizable 
interactions which respect the flavor symmetry. 
In other words, Yukawa couplings are generically functions 
of dynamical variables which carry the flavor charges and 
acquire non-zero VEVs around the scale $M_U$. 
Eventually, at low energies there appear effective Yukawa couplings 
as constants up to renormalization group (RG) effects. 
Thus the Froggatt-Nielsen mechanism
\cite{Frog} 
is at work for the interactions. 
Specifically, the effective Yukawa interactions for up-type quarks 
are of the form 
\begin{equation}
  M^{(q)}_{ij} \, Q_i U^c_j H_u 
\label{eqn:upqY}
\end{equation}
with 
\begin{equation}
  M^{(q)}_{ij} = c_{ij} \left( \frac{\langle X \rangle}
                              {M_S}\right)^{m_{ij}} 
           = c_{ij} \, x^{m_{ij}}, 
\label{eqn:Mij}
\end{equation}
where the subscripts $i$ and $j$ are the generation indices, 
and all of the constants $c_{ij}$ are of $O(1)$ with 
rank\,$c_{ij}=3$. 
The superfield $X$, which is singlet under the unification gauge 
group $G$ but carries a certain flavor charge, 
is an appropriate composite superfield with canonical 
normalization. 
The scale $M_S$ is the intrinsic scale of the underlying string 
theory and is supposed to lie around the reduced Planck scale. 
The ratio $x= \langle X \rangle /M_S$ is assumed to be slightly 
less than 1. 
The large hierarchy results from raising $x$ to large powers. 
Due to the flavor symmetry, the exponents $m_{ij}$ are determined 
as the flavor-charges of $Q_i$, $U^c_j$ and $H_u$. 
Consequently, it is possible for us to obtain Yukawa hierarchies.

In this paper we solve for the extra-particle mixings which 
are closely intertwined with Yukawa hierarchies. 
With the assumption $G \supset SU(2)_R$, 
we show that the non-trivial texture of the CKM matrix originates 
from the extra-particle mixings purely within the effective 
field theory. 
The observed structure of the CKM matrix of quarks is 
attributable to a large mixing between the down-type quarks and 
colored Higgses, 
both of which are $SU(2)_L$-singlets. 
On the other hand, the CKM matrix of leptons emerges as a 
consequence of the seesaw mechanism
\cite{Seesaw} 
with the hierarchical Majorana mass matrix for 
right-handed neutrinos. 
In previous papers
\cite{QCKM}\cite{LCKM} 
the present author and his collaborators 
explored the fermion spectra and the CKM matrix based on a 
$SU(6) \times SU(2)_R$ string-inspired model. 
In this paper we show that the main results in the previous papers 
can be derived within a more general framework. 
Furthermore, the distinct origins of the CKM matrices of quarks 
and leptons are ascertained. 
Since we ignore the phase factors of VEVs for matter fields, 
we do not discuss the $CP$-violation.

This paper is organized as follows. 
Assuming that the unification gauge group $G$ contains 
$SU(2)_R$ as well as $G_{SM}$, 
in \S 2 we present the mass matrices of quarks and leptons 
which arise from the Froggatt-Nielsen mechanism. 
These mass matrices, except for up-type quarks, 
imply the extra-particle mixings. 
After diagonalizing the mass matrices by bi-unitary 
transformations, 
we trace the origin of the CKM matrices for quarks and leptons. 
In \S 3 we introduce the flavor symmetry 
$U(1) \times {\bf Z}_2$. 
It is shown that the textures of the CKM matrices are 
controlled by the flavor charges, and so by the magnitude of 
the extra-particle mixings. 
In \S 4 we study the dependence of the relations 
among the fermion spectra, $V^Q_{\rm CKM}$ and $V^L_{\rm CKM}$, 
on the unification gauge group $G$. 
Section 5 is devoted to a summary.

\section{Mass matrices and their diagonalization}

Based on the $E_6$-type model, 
we assume that the unification gauge group $G$ is rank 6, 
and that $G$ contains $SU(2)_R$ as well as $G_{SM}$. 
The matter chiral superfields, if we represent them in terms of 
$E_6$, consist of 
\begin{equation} 
       N_f \,{\bf 27} \ 
               + \ \delta \,({\bf 27} + {\bf 27^*}). 
\end{equation} 
Here $N_f$ denotes the family number at low energies, 
and $\delta $ is the set number of vector-like multiplets. 
Note that adjoint or higher representation matter fields 
are not allowed in the level-one string theory. 
Let us now consider the case $N_f = 3$ and $\delta =1$. 
Matter fields in the ${\bf 27}$ representation of $E_6$ fall in 
two categories, 
\begin{eqnarray*} 
\Phi (SU(2)_R{\rm -singlet}) &:& \ \ Q, \ L, 
                                   \ g, \ g^c, \ S, \\
\Phi (SU(2)_R{\rm -doublet}) &:& \ \ (U^c, D^c), \ (N^c, E^c), 
                                    \ (H_u, H_d), 
\end{eqnarray*} 
where $g$, $g^c$ and $H_u$, $H_d$ represent colored Higgs 
and doublet Higgs fields, respectively. 
$N^c$ represents the right-handed neutrino superfield 
and $S$ is an $SO(10)$-singlet. 
When the gauge group $G$ becomes smaller than $E_6$ via flux 
breaking within the string theory, 
the vector-like multiplets potentially contain only some part of 
the ${\bf 27}$ representation but not all of it. 
The concrete matter contents of the vector-like multiplets are closely 
linked to the topological structure of the compactified space. 
Although $L$ and $H_d$ ($D^c$ and $g^c$) have the same quantum 
numbers under the standard model gauge group $G_{SM}$, 
they reside in different irreducible representations of $G$. 
Gauge invariant trilinear couplings are of the forms 
\begin{eqnarray} 
    \Phi ^3  &=&  QQg + Qg^cL + g^cgS + QH_dD^c + QH_uU^c 
                           + LH_dE^c          \nonumber \\ 
           {}& & \qquad  + LH_uN^c + SH_uH_d 
                          + gN^cD^c + gE^cU^c + g^cU^cD^c.
\end{eqnarray} 

We now introduce the ${\bf Z}_2$ symmtry($R$-parity) 
and attach odd $R$-parity to chiral multiplets 
$\Phi _i$ $(i=1,2,3)$ and even $R$-parity to one set of vector-like 
multiplets $\Phi _0$ and ${\overline \Phi}$. 
Ordinary quarks and leptons are included in chiral multiplets 
$\Phi _i$ $(i=1,2,3)$. 
Since light Higgs scalars are even under $R$-parity, 
light Higgs doublets reside in $\Phi _0$ and/or ${\overline \Phi}$.
The present scheme allows the $R$-parity to remain unbroken 
all the way down to the electroweak scale.
\cite{Majorana} 
Under an appropriate condition, the gauge symmetry $G$ 
is spontaneouly broken in two steps at the scales 
$\langle S_0 \rangle = \langle {\overline S \rangle }$ and 
$\langle N^c_0 \rangle = \langle {\overline N^c \rangle }$ as 
\begin{equation} 
   G \quad \buildrel \langle S_0 \rangle \over \longrightarrow 
   \quad G' \quad \buildrel \langle N^c_0 \rangle \over 
   \longrightarrow \quad G_{SM}=SU(3)_c \times SU(2)_L \times U(1)_Y, 
\end{equation} 
where $\langle S_0 \rangle = M_U$ and $G' \supset SU(2)_R$.
\cite{Aligned} 
In this model $SU(2)_R$ gauge symmtry is spontaneously 
broken at the scale $\langle N^c_0 \rangle$. 
In view of the phenomenological result that three gauge coupling 
constants of $G_{SM}$ meet around $2 \times 10^{16}$GeV in the MSSM, 
the scale $\langle N^c_0 \rangle$ is likely to be 
$O(10^{16}{\rm GeV})$.
\cite{Amaldi} 
Some particles could be lying somewhere between the scale $M_U$ 
and the TeV scale.

Let us discuss the mass matrices of quarks and leptons. 
To begin with the superpotential for up-type quarks is given by 
\begin{equation}
    W_U  \sim  X^{m_{ij}} Q_i U^c_j H_{u0}, 
\end{equation}
where $i,\ j = 1,\ 2,\ 3$ and all terms are characterized 
by couplings of $O(1)$ in units of $M_S$. 
The $G$-singlet superfield $X$ is assumed to carry a certain 
flavor charge. 
In the present framework the superfield $X$ is a holomorphic 
function of $S_0 {\overline S}$ and moduli fields. 
The exponents $m_{ij}$ are determined by the flavor symmetry. 
The $3 \times 3$ mass matrix $M^{(q)}_{ij} v_u$  of the up-type quarks 
can be diagonalized by a bi-unitary transformation as 
\begin{equation} 
    {\cal V}_u^{-1} M^{(q)} \,{\cal U}_u, 
\label{eqn:VMU}
\end{equation} 
where $v_u = \langle H_{u0} \rangle$. 
Hereafter all mass matrices are assumed to be non-singular. 
Although $x$ by itself is not a very small number, 
the physical parameters may be very small if they depend on 
high powers $m_{ij}$. 
We are not interested in obtaining the precise values of the 
parameters in the mass matrix but focus on orders of magnitude.

We next study the mass matrix for down-type quarks, 
in which there appear mixings between $g^c$ 
and $D^c$ at energies below the scale $\langle N_0^c \rangle$.
The superpotential of down-type colored fields is of the form 
\begin{equation}
    W_D  \sim  X^{z_{ij}} g_i g^c_j S_0
                    + X^{g_{ij}} g_i D^c_j N^c_0 
                          + X^{m_{ij}} Q_i D^c_j H_{d0}, 
\end{equation}
where the exponents $z_{ij}$, $g_{ij}$ and $m_{ij}$ are 
determined by flavor symmetry, and all terms are also 
multiplied by $O(1)$ couplings in $M_S$ units. 
Below $\langle N_0^c \rangle$ 
the mass matrix of down-type colored fields is expressed as 
the $6 \times 6$ matrix 
\begin{equation} 
\begin{array}{r@{}l} 
   \vphantom{\bigg(}   &  \begin{array}{ccc} 
          \quad   g^c   &  \quad \  D^c  &  
        \end{array}  \\ 
\widehat{M}_d = 
   \begin{array}{l} 
        g   \\  D  \\ 
   \end{array} 
     & 
\left( 
  \begin{array}{cc} 
       Z^{(q)}    &     G^{(q)}     \\
          0       &  \rho _d M^{(q)} 
  \end{array} 
\right) 
\end{array} 
\label{eqn:Mdh} 
\end{equation} 
in $M_S$ units,
\cite{QCKM} 
where we define 
\begin{equation}
    Z^{(q)}_{ij} = O(x^{z_{ij}}) 
             \frac {\langle S_0 \rangle}{M_S}, \qquad 
    G^{(q)}_{ij} = O(x^{g_{ij}}) 
             \frac {\langle N^c_0 \rangle}{M_S} 
\end{equation}
and $\rho _d = \langle H_{d0} \rangle /M_S = v_d /M_S$. 
$\widehat{M}_d$ yields mixings between $g^c$ and $D^c$ 
explicitly. 
An early attempt of explaining the CKM matrix via $g^c$-$D^c$ 
mixings appears in Ref.\cite{SO10}, 
in which a SUSY $SO(10)$ model is considered. 
Since $\rho _d$ is a very small number($\sim 10^{-16}$), 
it is plausible that the elements of the $3 \times 3$ submatrices 
$Z^{(q)}$ and $G^{(q)}$ are sufficiently large compared with those 
of $\rho _d M^{(q)}$. 
Therefore, the left-handed light quarks consist almost completely 
of $D$-components of the quark doublet $Q$. 
On the other hand, the mixing between $SU(2)_L$-singlet states 
$g^c$ and $D^c$ can be sizable, depending on the ratio of 
$G^{(q)}$ to $Z^{(q)}$. 
This type of mixing does not occur for up-type quarks. 
If all elements of $G^{(q)}$ are sufficiently small relative to 
those of $Z^{(q)}$, 
$\widehat{M}_d$ is approximately separated into two 
$3 \times 3$ submatrices. 
We have three heavy modes and three light modes which are 
mainly determined by $Z^{(q)}$ and $\rho _d M^{(q)}$, 
respectively. 
The three light modes correspond to $d$-, $s$- and $b$-quarks. 
In this case the mass spectra of the light down-type quarks is 
the same as those of the up-type quarks up to the ratio 
$v_u/v_d = \tan \beta$, 
and also $V^Q_{\rm CKM}$ becomes almost a unit matrix. 
This result is not preferable phenomenologically. 
Alternatively, if the elements of $G^{(q)}$ are comparable to or 
larger than those of $Z^{(q)}$, 
the mass spectra of down-type quarks are generally different from 
those of up-type quarks. 
In addition, we have a non-trivial CKM matrix. 
However, when the elements of $G^{(q)}$ are significantly larger 
than those of $Z^{(q)}$, 
the $d$-quark Yukawa coupling becomes too small relative to the 
$u$-quark coupling. 
Then we do not consider such a case. 
Thus, to obtain a phenomenologically viable solution, 
hereafter we consider a large mixing between $g^c$ and $D^c$, 
that is, $G^{(q)}_{ij} \sim Z^{(q)}_{ij}$.

$\widehat{M}_d$ can be diagonalized by a bi-unitary 
transformation as 
\begin{equation} 
    \widehat{\cal V}_d^{-1} \widehat{M}_d \, 
                        \widehat{\cal U}_d. 
\label{eqn:Md} 
\end{equation} 
Using a perturbative expansion with respect to $\rho _d$, 
we can solve the eigenvalue problem. 
After some algebra the $6 \times 6$ diagonalization unitary matrices 
$\widehat{\cal{V}}_d$ and $\widehat{\cal{U}}_d$ turn out to be 
\begin{eqnarray} 
   \widehat{\cal V}_d & \simeq & \left( 
   \begin{array}{cc} 
      {\cal W}_d   &  -\rho _d (A_d + B_d)^{-1} 
                              G^{(q)} M^{(q)\dag} {\cal V}_d \\
      \rho _d M^{(q)} G^{(q)\dag}(A_d + B_d)^{-1} {\cal W}_d    &  
                                          {\cal V}_d 
   \end{array} 
                       \right), 
\label{eqn:hatvd}                       \\ 
   \widehat{\cal U}_d & \simeq & \left( 
   \begin{array}{cc} 
      Z^{(q)\dag} {\cal W}_d \,(\Lambda _d^{(0)})^{-1} 
         &  -(Z^{(q)})^{-1} G^{(q)} (M^{(q)})^{-1}
                            {\cal V}_d \,\Lambda _d^{(2)}  \\
      G^{(q)\dag} {\cal W}_d \,(\Lambda _d^{(0)})^{-1} 
           &   (M^{(q)})^{-1} {\cal V}_d \,\Lambda _d^{(2)} 
   \end{array} 
   \right) 
\label{eqn:hatud}
\end{eqnarray} 
in the $\rho _d$ expansion, where 
\begin{equation} 
  A_d = Z^{(q)} Z^{(q)\dag}, \qquad B_d = G^{(q)} G^{(q)\dag}. 
\end{equation} 
The $3 \times 3$ unitary matrices ${\cal W}_d$ and ${\cal V}_d$ 
are determined such that the matrices 
\begin{eqnarray} 
  {} & & \qquad \qquad \qquad {\cal W}_d^{-1}(A_d + B_d){\cal W}_d 
                          = (\Lambda _d^{(0)})^2,   
\label{eqn:ABd}                                     \\
  {} & & {\cal V}_d^{-1} \left[ M^{(q)}(G^{(q)})^{-1}
       (A_d^{-1} + B_d^{-1})^{-1}(G^{(q)\dag})^{-1}M^{(q)\dag} \right] 
                    {\cal V}_d = (\Lambda _d^{(2)})^2 
\label{eqn:ABdI}
\end{eqnarray} 
become diagonal. 
The diagonal elements of $M_S \Lambda _d^{(0)}$ represent 
masses of three heavy modes which are expected to be roughly 
of $O(M_U)$. 
On the other hand, three diagonal elements of 
$v_d \Lambda _d^{(2)}$ represent masses of light modes 
corresponding to $d$-, $s$- and $b$-quarks. 
From Eq.(\ref{eqn:hatud}) the mass eigenstates of light 
$SU(2)_L$-singlet down-type quarks are given by 
\begin{equation}
   \tilde{D}^c \simeq \Lambda _d^{(2)} {\cal V}_d^{-1}
             (M^{(q)\dag})^{-1} G^{(q)\dag} 
             \left[ -(Z^{(q)\dag})^{-1} g^c + 
             (G^{(q)\dag})^{-1} D^c \right]. 
\end{equation}
As seen in Eq.(\ref{eqn:ABdI}), 
when the mixings between $g^c$ and $D^c$ are large, 
${\cal V}_d$ evidently differs from ${\cal V}_u$. 
Therefore, we obtain a nontrivial CKM matrix,
\cite{QCKM} 
\begin{equation} 
   V^Q_{\rm CKM} = {\cal V}_u^{-1} \,{\cal V}_d. 
\end{equation} 

We next proceed to study the mass matrices for the lepton sector. 
Under $G_{SM}$, the colorless $SU(2)_L$-doublet fields $L$ and $H_d$ 
are not distinguished from each other, 
and then $L$-$H_d$ mixing occurs at energies below the scale 
$\langle N_0^c \rangle$. 
Since both $L$ and $H_d$ are $SU(2)_L$-doublets, 
$L$-$H_d$ mixing does not trigger the disparity between 
diagonalization matrices for the up-sector and the down-sector 
of leptons. 
However, the disparity is possibly triggered by another mixing, 
namely, by the seesaw mechanism for neutrinos. 
For charged leptons the superpotential is 
\begin{equation} 
   W_E  \sim  X^{h_{ij}} H_{di} H_{uj} S_0 
           + X^{k_{ij}} L_i H_{uj} N^c_0 
           + X^{l_{ij}} L_i E^c_j H_{d0} 
\end{equation} 
in $M_S$ units, 
where the exponents $h_{ij}$, $k_{ij}$ and $l_{ij}$ are 
determined by the flavor symmetry. 
Below the scale $\langle N^c_0 \rangle$, the mass matrix for 
charged leptons has the form 
\begin{equation} 
\begin{array}{r@{}l} 
   \vphantom{\bigg(}   &  \begin{array}{ccc} 
          \ \  H_u^+   &  \quad E^{c+}  &  
        \end{array}  \\ 
\widehat{M}_l = 
   \begin{array}{l} 
        H_d^-  \\  L^-  \\ 
   \end{array} 
     & 
\left( 
  \begin{array}{cc} 
      Z^{(l)}   &        0         \\
      G^{(l)}   &  \rho _d M^{(l)}  
  \end{array} 
\right) 
\end{array} 
\label{eqn:Mlh} 
\end{equation} 
in $M_S$ units, 
where 
\begin{equation}
   Z^{(l)}_{ij} = O(x^{h_{ij}}) 
             \frac {\langle S_0 \rangle}{M_S}, \qquad 
   G^{(l)}_{ij} = O(x^{k_{ij}})
             \frac {\langle N^c_0 \rangle}{M_S}, \qquad 
   M^{(l)}_{ij} = O(x^{l_{ij}}). 
\end{equation}
Note that the matrix $Z^{(l)}$ is symmetric. 
This $6 \times 6$ matrix $\widehat{M}_l$ exhibits a texture quite 
similar to that of $\widehat{M}_d^{\dag}$. 
Then, by observing the one-to-one correspondence 
$Z^{(l)} \leftrightarrow Z^{(q)\dag}$, 
$G^{(l)} \leftrightarrow G^{(q)\dag}$ and 
$M^{(l)} \leftrightarrow M^{(q)\dag}$, 
the diagonalization of $\widehat{M}_l$ is closely analogous 
to that in the case of down-type quarks. 
In fact, $\widehat{M}_l$ is diagonalized by 
the bi-unitary transformation 
\begin{equation} 
    \widehat{\cal V}_l^{-1} \widehat{M}_l \, 
                        \widehat{\cal U}_l 
\label{eqn:Ml} 
\end{equation} 
with 
\begin{eqnarray} 
   \widehat{\cal V}_l & \simeq & \left( 
   \begin{array}{cc} 
      Z^{(l)} {\cal W}_l \,(\Lambda _l^{(0)})^{-1}  
         &  -(Z^{(l)\dag})^{-1} G^{(l)\dag} 
                (M^{(l)\dag})^{-1}{\cal V}_l 
                          \,\Lambda _l^{(2)}  \\
      G^{(l)} {\cal W}_l \,(\Lambda _l^{(0)})^{-1}   
         &  (M^{(l)\dag})^{-1} {\cal V}_l 
                         \,\Lambda _l^{(2)} \\
   \end{array} 
                       \right), 
\label{eqn:hatvl}                              \\
   \widehat{\cal U}_l & \simeq & \left( 
   \begin{array}{cc} 
      {\cal W}_l   &  -\rho _d (A_l + B_l)^{-1} 
                          G^{(l)\dag} M^{(l)} {\cal V}_l \\
      \rho _d M^{(l)\dag} G^{(l)} (A_l + B_l)^{-1} {\cal W}_l  & 
                                           {\cal V}_l 
   \end{array} 
                       \right) 
\label{eqn:hatul}
\end{eqnarray} 
in the $\rho _d$ expansion, 
where 
\begin{equation}
   A_l = Z^{(l)\dag} Z^{(l)}, \qquad 
   B_l = G^{(l)\dag}G^{(l)}. 
\end{equation}
${\cal W}_l$ and ${\cal V}_l$ are unitary matrices such that 
\begin{eqnarray} 
  {} & & \qquad \qquad \qquad {\cal W}_l^{-1} (A_l + B_l) {\cal W}_l 
                     = (\Lambda _l^{(0)})^2,   
\label{eqn:Wl}                                     \\
  {} & & {\cal V}_l^{-1} \left[ M^{(l)\dag} (G^{(l)\dag})^{-1} 
       (A_l^{-1} + B_l^{-1})^{-1} (G^{(l)})^{-1} M^{(l)} \right] 
                 {\cal V}_l = (\Lambda _l^{(2)})^2 
\label{eqn:Vl} 
\end{eqnarray} 
are diagonal. 
The three diagonal elements of $v_d \Lambda ^{(2)}_l$ represent masses of 
light charged leptons corresponding to $e$, $\mu $ and $\tau $. 
Equation (\ref{eqn:hatul}) implies that 
the light $SU(2)_L$-singlet charged leptons are 
mainly $E^c$-components. 
From Eq.(\ref{eqn:hatvl}), the mass eigenstates of 
light $SU(2)_L$-doublet charged leptons are given by 
\begin{equation}
    \tilde{L}^{-} \simeq \Lambda _l^{(2)} {\cal V}_l^{-1} (M^{(l)})^{-1} G^{(l)} 
        \left[ - (Z^{(l)})^{-1} H_d^{-} + (G^{(l)})^{-1} L^{-} \right]. 
\label{eqn:l-}
\end{equation}
Consequently, when the elements of $(Z^{(l)})^{-1}$ and 
$(G^{(l)})^{-1}$ are comparable to each other, 
there occurs a large mixing between $H_d^-$ and $L^-$.

In the neutral lepton sector we have the superpotential 
\begin{eqnarray} 
   W_N  & \sim &  X^{h_{ij}} H_{di} H_{uj} S_0 
                + X^{k_{ij}} L_i H_{uj} N^c_0 
                + X^{l_{ij}} L_i N^c_j H_{u0} \nonumber \\
     {} & &  \qquad  + X^{n_{ij}} (N^c_i {\overline N}^c) 
                              (N^c_j {\overline N}^c) 
\end{eqnarray} 
in $M_S$ units. 
Below $\langle N^c_0 \rangle$ the mass matrix  for neutral leptons 
becomes the $12 \times 12$ matrix 
\begin{equation} 
\begin{array}{r@{}l} 
   \vphantom{\bigg(}   &  \begin{array}{ccccc} 
          \quad  H_u^0   & \  H_d^0  &  \quad  L^0  
                          &  \qquad  N^c   &
        \end{array}  \\ 
\widehat{M}_N = 
   \begin{array}{l} 
        H_u^0  \\  H_d^0  \\  L^0  \\  N^c  \\
   \end{array} 
     & 
\left( 
  \begin{array}{cccc} 
        0    & Z^{(l)} &   G^{(l)T}    &      0          \\
     Z^{(l)} &    0    &     0         &      0          \\
     G^{(l)} &    0    &     0         & \rho _u M^{(l)} \\
        0    &    0    & \rho _u M^{(l)T} &      N       \\
  \end{array} 
\right) 
\end{array} 
\label{eqn:Mn} 
\end{equation} 
in $M_S$ units, where 
\begin{equation}
   N_{ij} = O(x^{n_{ij}}) 
          \left( \frac{\langle N^c_0 \rangle}{M_S} \right)^2 
\end{equation}
and $\rho _u = v_u/M_S$. 
More precisely, we have to take the contribution of the field $S$ 
into account. 
However, to illustrate the general features, 
the above $12 \times 12$ mass matrix suffices for us, 
because the main result remains unchanged. 
By recalling the above study in the charged lepton sector, 
we first carry out the unitary transformation 
\begin{equation}
   \widehat{\cal U}_P^{-1} \widehat{M}_N \widehat{\cal U}_P 
\end{equation}
with 
\begin{equation}
  \widehat{\cal U}_P = 
   \left( 
   \begin{array}{ccc}
      {\cal W}_l  &      0      &      0       \\
           0      &  \widehat{\cal V}_l  &  0  \\
           0      &      0      &  {\cal U}_N   
   \end{array}
   \right). 
\end{equation}
Note that $\widehat{\cal V}_l$ is the $6 \times 6$ matrix 
given by Eq.(\ref{eqn:hatvl}), 
and ${\cal W}_l$ is defined in Eq.(\ref{eqn:Wl}). 
The $3 \times 3$ unitary matrix ${\cal U}_N$ diagonalizes the Majorana 
mass matrix $N$ as 
\begin{equation}
   {\cal U}_N^{-1} N {\cal U}_N = \Lambda _N, 
\end{equation}
where $\Lambda _N$ is diagonal. 
The scales of $M_S\,\Lambda_N$ are assumed to lie in the range 
$10^{10} \sim 10^{12}$GeV. 
As a result, we obtain an approximately separated matrix which is 
composed of the $6 \times 6$ submatrix for heavy modes and 
the seesaw-type $6 \times 6$ submatrix. 
The submatrix for heavy modes leads to the same masses as the heavy modes 
of charged leptons. 
The seesaw-type submatrix is of the form 
\begin{equation}
   \left( 
   \begin{array}{cc}
        0   &   \rho _u \,\Lambda ^{(2)}_l {\cal V}^{-1}_l {\cal U}_N   \\
      \rho _u \,{\cal U}_N^{-1} {\cal V}_l \Lambda ^{(2)}_l   
                               &   {\cal U}_N^{-1} N {\cal U}_N   
   \end{array}
   \right). 
\label{eqn:sss}
\end{equation}
Thus $\widehat{M}_N$ can be diagonalized as 
\begin{equation}
   \widehat{\cal U}_N^{-1} \widehat{M}_N \widehat{\cal U}_N 
\end{equation}
with 
\begin{equation}
   \widehat{\cal U}_N = 
               \widehat{\cal U}_P \times \widehat{\cal U}_Q 
\label{eqn:tilun}
\end{equation}
and 
\begin{equation}
  \widehat{\cal U}_Q \simeq 
   \left( 
   \begin{array}{cccc}
      1/\sqrt 2 & -1/\sqrt 2 &     0     &   0   \\
      1/\sqrt 2 &  1/\sqrt 2 &     0     &   0   \\
         0      &      0     &  {\cal V} 
      &   \rho _u \,\Lambda ^{(2)}_l {\cal V}^{-1}_l N^{-1} {\cal U}_N    \\
         0      &      0     &  
      - \rho _u \,{\cal U}_N^{-1} N^{-1} {\cal V}_l \Lambda ^{(2)}_l {\cal V} 
                                         &   1   
   \end{array}
   \right) 
\label{eqn:UQ}
\end{equation}
in the $\rho _u$ expansion. 
The unitary matrix ${\cal V}$ in Eq.(\ref{eqn:UQ}) is defined 
such that 
\begin{equation}
     {\cal V}^{-1} {\cal N} {\cal V} 
\label{eqn:VLVN}
\end{equation}
is diagonal. 
From Eq.(\ref{eqn:sss}) the mass matrix of light neutrinos has the form 
\begin{equation}
     \rho _u^2 \,{\cal N} = \rho _u^2 \,\Lambda ^{(2)}_l 
              {\cal V}^{-1}_l N^{-1} {\cal V}_l \Lambda ^{(2)}_l. 
\label{eqn:calN}
\end{equation}
Masses of light neutrinos become 
\begin{equation}
   m_{\nu } \simeq \frac {v_u^2}{M_S} 
               (\Lambda ^{(2)}_l)^2 \Lambda _N^{-1}. 
\end{equation}
It turns out that the light neutrino mass eigenstates 
are 
\begin{equation}
     \tilde{L}^0 \simeq 
    {\cal V}^{-1} \Lambda _l^{(2)} {\cal V}_l^{-1} (M^{(l)})^{-1} G^{(l)}
           \left[ - (Z^{(l)})^{-1} H_d^0 + (G^{(l)})^{-1} L^0 \right]. 
\label{eqn:l0}
\end{equation}
Comparing these eigenstates $\tilde{L}^0$ with those of 
the light charged leptons $\tilde{L}^-$ given 
by Eq.(\ref{eqn:l-}), 
we find that ${\cal V}$ is nothing but the CKM matrix for leptons, 
that is, 
\begin{equation}
   V^L_{\rm CKM} = {\cal V}. 
\end{equation}
As seen in Eq.(\ref{eqn:calN}), if the Majorana mass matrix $N$ is 
proportional to the unit matrix, 
we find that $V^L_{\rm CKM} = 1$. 
Contrastively, when $N$ contains hierarchical structure, 
we have a non-trivial CKM matrix for leptons. 
In the next section we explore such a case.

\section{The flavor symmetry and mass hierarchies}

The superpotential above the scale $M_U$ respects 
flavor symmetry as well as the gauge symmetry $G$. 
We now introduce the global $U(1) \times {\bf Z}_2$ 
as the flavor symmetry. 
As mentioned above, the ${\bf Z}_2$-charge is identified with 
$R$-parity in the MSSM. 
The charge of the $G$-singlet superfield $X$ is assumed 
to be $(-1,\ +)$ under $U(1) \times {\bf Z}_2$. 
Due to the $U(1)$ symmetry, 
the exponents $m_{ij}$, $z_{ij}$, {\it etc}., 
are determined as the $U(1)$-charges of matter 
superfields. 
Denote the $U(1)$-charges of the $i$-th generation matter fields 
$\Phi _{Ai}$ as $a_{Ai}$ $(i=1,2,3)$, 
where the $\Phi _A$ in the ${\bf 27}$ representation of $E_6$ are 
classified as 
\begin{equation}
   \Phi _A = \left\{  \begin{array}{lr}
              Q      &  \qquad \mbox{for $A = 1$}, \\
         (U^c,\ D^c) &  \qquad \mbox{for $A = 2$}, \\
              g      &  \qquad \mbox{for $A = 3$}, \\
             g^c     &  \qquad \mbox{for $A = 4$}, \\
              L      &  \qquad \mbox{for $A = 5$}, \\
         (N^c,\ E^c) &  \qquad \mbox{for $A = 6$}, \\
         (H_u,\ H_d) &  \qquad \mbox{for $A = 7$}, \\
              S      &  \qquad \mbox{for $A = 8$}.  
                      \end{array}
             \right.
\label{eqn:PhiA}
\end{equation}
Hereafter we define $\alpha _A$ and $\gamma _A$ by 
\begin{equation}
  a_{A2} - a_{A1} = \alpha _A, \qquad 
  a_{A3} - a_{A2} = \gamma _A. 
\end{equation}
All of the $\alpha _A$ and $\gamma _A$ are assumed to be positive. 
Thus the exponents which appear in the superpotential, 
for instance the $m_{ij}$, are of the form 
\begin{equation} 
   m_{ij} -m_{33} = \left(
      \begin{array}{ccc}
        \alpha _1 + \gamma _1 + \alpha _2 + \gamma _2  &  
                \alpha _1 + \gamma _1 + \gamma _2  &  
                      \alpha _1 + \gamma _1   \\ 
        \gamma _1 + \alpha _2 + \gamma _2  &  
              \gamma _1 + \gamma _2  &  \gamma _1  \\ 
        \alpha _2 + \gamma _2  &  \gamma _2   &  0   \\
      \end{array} 
             \right)_{ij}, 
\end{equation} 
provided that $m_{33}$ is non-negative. 
If we define the diagonal matrices according to 
\begin{equation} 
   \Gamma _A = \left(
      \begin{array}{ccc}
        x^{\alpha _A + \gamma _A}  &  0   &   0   \\ 
        0    &  x^{\gamma _A}  &  0  \\ 
        0    &     0   &  1   \\
      \end{array} 
             \right), \qquad (A=1,2, \cdots, 8) 
\end{equation} 
then the $3 \times 3$ mass matrix $M^{(q)}$ for up-type quarks 
can be rewritten in the factorized form 
\begin{equation}
    M^{(q)} =  \Gamma _1 \left[ \,y^{(q)}_M \,M^{(q)}_0 \right] \Gamma _2, 
\label{eqn:Mqgg}
\end{equation}
where $M^{(q)}_{0ij} = O(1)$ and $y^{(q)}_M = x^{m_{33}}$. 
Similarly, the other $3 \times 3$ submatrices can be expressed in 
the factorized forms 
\begin{eqnarray}
   {} & & \, Z^{(q)}  =  \Gamma _3 \left[ \,y^{(q)}_Z \,
                       Z^{(q)}_0 \right] \Gamma _4,   \qquad 
     G^{(q)}  =  \Gamma _3 \left[ \,y^{(q)}_G \,
                       G^{(q)}_0 \right] \Gamma _2,  \nonumber \\ 
   {} & & \, Z^{(l)}  =  \Gamma _7 \left[ \,y^{(l)}_Z \,
                       Z^{(l)}_0 \right] \Gamma _7,   \qquad \ 
     G^{(l)}  =  \Gamma _5 \left[ \,y^{(l)}_G \,
                       G^{(l)}_0 \right] \Gamma _7,  
\label{eqn:GG}                                 \\ 
   {} & & M^{(l)}  =  \Gamma _5 \left[ \,y^{(l)}_M \,
                       M^{(l)}_0 \right] \Gamma _6,   \qquad 
    \ \  N  =  \Gamma _6 \left[ \,y_N \,N_0 \,\right] \Gamma _6, 
                                 \nonumber 
\end{eqnarray}
where all elements of $Z^{(q)}_0$, $G^{(q)}_0$, $Z^{(l)}_0$, 
$G^{(l)}_0$, $M^{(l)}_0$ and $N_0$ are of $O(1)$. 
The constant factors are 
\begin{eqnarray}
   {} & & y^{(q)}_Z = x^{z_{33}} 
           \frac {\langle S_0 \rangle}{M_S}, \qquad 
   y^{(q)}_G = x^{g_{33}} \frac {\langle N^c_0 \rangle}{M_S}, 
                                       \nonumber  \\
   {} & & y^{(l)}_Z = x^{h_{33}} 
           \frac {\langle S_0 \rangle}{M_S}, \qquad 
   y^{(l)}_G = x^{k_{33}} \frac {\langle N^c_0 \rangle}{M_S}, \\
   {} & & y^{(l)}_M = x^{l_{33}},  \qquad \qquad \ 
   y_N = x^{n_{33}} 
            \left( \frac {\langle N^c_0 \rangle}{M_S} \right)^2, 
                                       \nonumber 
\label{eqn:yy}
\end{eqnarray}
provided that $z_{33}, \ g_{33}, \ h_{33}, \ k_{33}, 
\ l_{33}, \ n_{33} \geq 0$. 
It should be noted that due to the flavor symmetry, 
each mass matrix is given by a product of two $\Gamma _A$ and 
a matrix between them, 
whose elements are of the same order of magnitude. 
Therefore, the texture of the diagonalization matrices 
is essentially determined by those of these two $\Gamma _A$.

In the case of up-type quarks, from Eqs.(\ref{eqn:VMU}) and 
(\ref{eqn:Mqgg}), the texture of the diagonalization matrices 
${\cal V}_u$ and ${\cal U}_u$ of $M^{(q)}$ is governed by 
$\Gamma _1$ and $\Gamma _2$, respectively. 
Explicitly, ${\cal V}_u$ and ${\cal U}_u$ become 
\begin{eqnarray} 
    {\cal V}_u & = & 
      \left( 
      \begin{array}{ccc}
        1 - O(x^{2\alpha _1})  &  O(x^{\alpha _1})  
                          &  O(x^{\alpha _1+ \gamma _1})  \\
        O(x^{\alpha _1})    &  1 - O(x^{2\alpha _1})  
                                   &  O(x^{\gamma _1})  \\
        O(x^{\alpha _1 + \gamma _1})   &  O(x^{\gamma _1})  
                              &  1 - O(x^{2\gamma _1})  \\
      \end{array}
      \right), \\
    {\cal U}_u & = & 
      \left( 
      \begin{array}{ccc}
        1 - O(x^{2\alpha _2})  &  O(x^{\alpha _2})  
                          &  O(x^{\alpha _2 + \gamma _2 })  \\
        O(x^{\alpha _2 })    &  1 - O(x^{2\alpha _2 })  
                                   &  O(x^{\gamma _2})  \\
        O(x^{\alpha _2 + \gamma _2})   &  O(x^{\gamma _2})  
                              &  1 - O(x^{2\gamma _2})  \\
      \end{array}
      \right). 
\end{eqnarray} 
The diagonalization of $M^{(q)}$ yields the eigenvalues 
\begin{equation}
   x^{m_{33}} \times \left( 
          O(x^{\alpha _1 + \gamma _1 +\alpha _2 + \gamma _2}), \quad 
          O(x^{\gamma _1 + \gamma _2}), \quad 
          O(1)   \right). 
\end{equation}
We now assume that the global $U(1)$ charges are assigned 
in such a way that trilinear couplings of up-type quarks are allowed 
only for the third generation, i.e. 
\begin{equation}
   m_{33} = 0. 
\label{eqn:m33}
\end{equation}
It follows that the mass eigenvalues are 
\begin{equation}
   v_u \times \left( 
   O(x^{\alpha _1 + \gamma _1 +\alpha _2 + \gamma _2}), \qquad 
   O(x^{\gamma _1 + \gamma _2}), \qquad  
   O(1) \right), 
\end{equation}
which correspond to $u$-, $c$- and $t$-quarks, respectively.

For down-type quarks, the diagonalization matrices 
$\widehat{\cal V}_d$ and $\widehat{\cal U}_d$ of $\widehat{M}_d$, 
which are given by Eqs. (\ref{eqn:hatvd}) and (\ref{eqn:hatud}), 
are expressed in terms of ${\cal W}_d$ and ${\cal V}_d$. 
Here ${\cal W}_d$ and ${\cal V}_d$ are the diagonalization matrices 
for 
\begin{eqnarray}
   {} & & A_d + B_d = \Gamma _3 \left[ 
                     {\cal A}_d + {\cal B}_d \right] \Gamma _3, \\ 
   {} & & M^{(q)} \left( G^{(q)} \right)^{-1}
                 (A_d^{-1} + B_d^{-1})^{-1} 
       \left( G^{(q)\dag} \right)^{-1}  M^{(q)\dag} 
       = \Gamma _1 \left[ {\cal C}_d + {\cal D}_d 
        \right]^{-1} \Gamma _1, 
\label{eqn:ABdG}
\end{eqnarray}
where 
\begin{eqnarray}
    {\cal A}_d & = & \left( y^{(q)}_Z \right)^2 
                          Z^{(q)}_0 {\Gamma _4}^2 Z^{(q)\dag}_0,  \\
    {\cal B}_d & = & \left( y^{(q)}_G \right)^2 
                          G^{(q)}_0 {\Gamma _2}^2 G^{(q)\dag}_0,  \\
    {\cal C}_d & = & \left( \frac {y^{(q)}_G}{y^{(q)}_Z} \right)^2 
              ((Z^{(q)}_0)^{-1} G^{(q)}_0 (M^{(q)}_0)^{-1})^{\dag} 
               {\Gamma _4}^{-2} 
                ((Z^{(q)}_0)^{-1} G^{(q)}_0 (M^{(q)}_0)^{-1}), \\
    {\cal D}_d & = & 
           (M^{(q)\dag}_0)^{-1} {\Gamma _2}^{-2} (M^{(q)}_0)^{-1}. 
\end{eqnarray}
By observing that elements of the matrices ${\cal A}_d$, ${\cal B}_d$, 
${\cal C}_d$ and ${\cal D}_d$ are 
\begin{eqnarray}
    {\cal A}_{dij} & = & ( y^{(q)}_Z )^2 \times O(1),  \\
    {\cal B}_{dij} & = & ( y^{(q)}_G )^2 \times O(1),  \\
    {\cal C}_{dij} & = & \left( \frac {y^{(q)}_G}{y^{(q)}_Z} \right)^2 
                    \times O(x^{- 2 \alpha _4 - 2 \gamma _4}),  \\
    {\cal D}_{dij} & = & O(x^{- 2 \alpha _2 - 2 \gamma _2}),   
\end{eqnarray}
it is easy to obtain the diagonalization matrices 
\begin{eqnarray} 
    {\cal W}_d & = & 
      \left( 
      \begin{array}{ccc}
        1 - O(x^{2\alpha _3})  &  O(x^{\alpha _3})  
                          &  O(x^{\alpha _3+ \gamma _3})  \\
        O(x^{\alpha _3})    &  1 - O(x^{2\alpha _3})  
                                   &  O(x^{\gamma _3})  \\
        O(x^{\alpha _3 + \gamma _3})   &  O(x^{\gamma _3})  
                              &  1 - O(x^{2\gamma _3})  \\
      \end{array}
      \right), \\
    {\cal V}_d & = & 
      \left( 
      \begin{array}{ccc}
        1 - O(x^{2\alpha _1})  &  O(x^{\alpha _1})  
                          &  O(x^{\alpha _1 + \gamma _1 })  \\
        O(x^{\alpha _1 })    &  1 - O(x^{2\alpha _1 })  
                                   &  O(x^{\gamma _1})  \\
        O(x^{\alpha _1 + \gamma _1})   &  O(x^{\gamma _1})  
                              &  1 - O(x^{2\gamma _1})  \\
      \end{array}
      \right). 
\end{eqnarray} 
Note that corresponding elements of the matrices ${\cal V}_u$ 
and ${\cal V}_d$ are of the same order of magnitude. 
To get a phenomenologically viable solution, 
we assume that a large mixing occurs between $g^c$ and $D^c$. 
This situation is realized by assuming 
$(Z^{(q)})_{11} \sim (G^{(q)})_{11}$; namely, 
\begin{equation}
    y^{(q)}_Z \, x^{\alpha _4 + \gamma _4} \simeq 
                y^{(q)}_G \, x^{\alpha _2 + \gamma _2}. 
\end{equation}
In this case the two terms ${\cal C}_d$ and ${\cal D}_d$ in the 
brackets of Eq.(\ref{eqn:ABdG}) become comparable, 
and so coefficients of the leading terms in off-diagonal elements 
of ${\cal V}_u$ and ${\cal V}_d$ are different. 
Consequently, the CKM matrix is given by
\cite{QCKM} 
\begin{equation}
  V^Q_{\rm CKM} = {\cal V}_u^{-1} {\cal V}_d 
         = \left( 
      \begin{array}{ccc}
        1 - O(x^{2\alpha _1})  &  O(x^{\alpha _1})  
                          &  O(x^{\alpha _1 + \gamma _1 })  \\
        O(x^{\alpha _1 })    &  1 - O(x^{2\alpha _1 })  
                                   &  O(x^{\gamma _1})  \\
        O(x^{\alpha _1 + \gamma _1})   &  O(x^{\gamma _1})  
                              &  1 - O(x^{2\gamma _1})  \\
      \end{array}
      \right). 
\label{eqn:QCKM}
\end{equation} 
As pointed out in the previous section, 
when $(Z^{(q)})_{11} \gg (G^{(q)})_{11}$, 
the CKM matrix becomes almost a unit matrix. 
It is worth noting that large $g^c$-$D^c$ mixings play 
an essential role in generating a non-trivial texture of 
the CKM matrix. 
Furthermore, the mass eigenvalues of light down-type quarks, 
which are expressed as $\Lambda ^{(2)}_d$ in Eq.(\ref{eqn:ABdI}), 
turn out to be 
\begin{equation}
   v_d \times \left( 
   O(x^{\alpha _1 + \gamma _1 +\alpha _2 + \gamma _2}), \quad 
   O(x^{\gamma _1 + \alpha _2 + \gamma _2}), \quad 
   O(x^{\alpha _2 + \gamma _2 - \delta _d}) \right), 
\end{equation}
where 
\begin{equation}
    \delta _d= \min ( \alpha _2, \ \alpha _4 ). 
\end{equation}
Note that $\alpha _2$ and $\alpha _4$ represent the differences 
between $U(1)$ charges, $a_{A2} - a_{A1}$, for $A = D^c$ and 
$A = g^c$, respectively. 
By taking $\gamma _1 = 2\alpha _1$ and 
$x^{\alpha _1} = \lambda \simeq 0.22$ in Eq.(\ref{eqn:QCKM}), 
we obtain the well-known form of the CKM matrix, 
\begin{equation} 
   V_{\rm CKM}^Q \sim \left(
      \begin{array}{ccc}
        1 - O(\lambda ^2)  &  O(\lambda )  &  
                              O(\lambda ^3)   \\ 
        O(\lambda )  &  1 - O(\lambda ^2)  &  
                              O(\lambda ^2)   \\ 
        O(\lambda ^3)  &   O(\lambda ^2) & 
                              1 - O(\lambda ^4)  
      \end{array} 
             \right), 
\label{eqn:QCKMN} 
\end{equation} 
up to RG effects. 
In addition, if we take $\alpha _2 \sim 2.5\alpha _1$ and 
$\gamma _2 \sim 1.5\alpha _1$, 
the quark spectra are also reproduced approximately.

For the lepton sector the diagonalization matrices 
$\widehat{\cal V}_l$ and $\widehat{\cal U}_l$ of $\widehat{M}_l$ 
given by Eqs. (\ref{eqn:hatvl}) and (\ref{eqn:hatul}) are described 
in terms of ${\cal W}_l$ and ${\cal V}_l$, 
which are the diagonalization matrices for 
\begin{eqnarray}
   {} & & A_l + B_l = \Gamma _7 \left[ 
               {\cal A}_l + {\cal B}_l \right] \Gamma _7, 
\label{eqn:ABlG1}                \\ 
   {} & & M^{(l)\dag} \left( G^{(l)\dag} \right)^{-1}
                 (A_l^{-1} + B_l^{-1})^{-1} 
       \left( G^{(l)} \right)^{-1}  M^{(l)} 
       = \Gamma _6 \left[ {\cal C}_l + {\cal D}_l 
        \right]^{-1} \Gamma _6, 
\label{eqn:ABlG2}
\end{eqnarray}
where 
\begin{eqnarray}
    {\cal A}_l & = & \left( y^{(l)}_Z \right)^2 
                          Z^{(l)\dag}_0 {\Gamma _7}^2 Z^{(l)}_0,  \\
    {\cal B}_l & = & \left( y^{(l)}_G \right)^2 
                          G^{(l)\dag}_0 {\Gamma _5}^2 G^{(l)}_0,  \\
    {\cal C}_l & = & \left( \frac {y^{(l)}_G}
                             {y^{(l)}_M \,y^{(l)}_Z} \right)^2 
              ((M^{(l)}_0)^{-1} G^{(l)}_0 (Z^{(l)}_0)^{-1}) 
               {\Gamma _7}^{-2} 
                ((M^{(l)}_0)^{-1} G^{(l)}_0 (Z^{(l)}_0)^{-1})^{\dag}, \\
    {\cal D}_l & = & (y^{(l)}_M)^{-2} 
           (M^{(l)}_0)^{-1} {\Gamma _5}^{-2} (M^{(l)\dag}_0)^{-1}. 
\end{eqnarray}
Equations (\ref{eqn:ABlG1}) and (\ref{eqn:ABlG2}) lead to 
\begin{eqnarray} 
    {\cal W}_l & = & 
      \left( 
      \begin{array}{ccc}
        1 - O(x^{2\alpha _7})  &  O(x^{\alpha _7})  
                          &  O(x^{\alpha _7+ \gamma _7})  \\
        O(x^{\alpha _7})    &  1 - O(x^{2\alpha _7})  
                                   &  O(x^{\gamma _7})  \\
        O(x^{\alpha _7 + \gamma _7})   &  O(x^{\gamma _7})  
                              &  1 - O(x^{2\gamma _7})  \\
      \end{array}
      \right), \\
    {\cal V}_l & = & 
      \left( 
      \begin{array}{ccc}
        1 - O(x^{2\alpha _6})  &  O(x^{\alpha _6})  
                          &  O(x^{\alpha _6 + \gamma _6 })  \\
        O(x^{\alpha _6 })    &  1 - O(x^{2\alpha _6 })  
                                   &  O(x^{\gamma _6})  \\
        O(x^{\alpha _6 + \gamma _6})   &  O(x^{\gamma _6})  
                              &  1 - O(x^{2\gamma _6})  \\
      \end{array}
      \right). 
\end{eqnarray} 
To obtain phenomenologically viable lepton spectra, 
we now assume $\alpha _7 > \xi > 0$, 
where $\xi $ is defined by 
\begin{equation}
    x^{\xi } = \frac {y^{(l)}_Z}{ y^{(l)}_G} 
     \, x^{\alpha _7 + \gamma _7 - \alpha _5 - \gamma _5}. 
\end{equation}
This implies that a large mixing occurs between $L$ and $H_d$, 
and that ${\cal C}_l$ and ${\cal D}_l$ in the brackets of 
Eq.(\ref{eqn:ABlG2}) become comparable. 
The mass eigenvalues of light charged leptons, 
which are given by $\Lambda ^{(2)}_l$ in Eq.(\ref{eqn:Vl}), are 
\begin{equation}
   v_d \,y^{(l)}_M \times \left( 
   O(x^{\alpha _6 + \gamma _6 
                     + \alpha _5 + \gamma _5 + \xi}), \quad 
   O(x^{\gamma _6 + \alpha _5 + \gamma _5}), \quad 
   O(x^{\alpha _5 + \gamma _5 - \delta _l}) \right), 
\end{equation}
where 
\begin{equation}
    \delta _l= \min ( \alpha _5, \ \alpha _7 - \xi). 
\label{eqn:dell}
\end{equation}
Note that $\alpha _5$ and $\alpha _7$ represent the differences 
between $U(1)$ charges, $a_{A2} - a_{A1}$, for $A = L$ and 
$A = H_d$, respectively. 
Since both $L$ and $H_d$ are $SU(2)_L$-doublets, 
a non-trivial CKM matrix for leptons is not derived as a result 
of the $L$-$H_d$ mixing but arises from the seesaw mechanism. 
As given in Eq.(\ref{eqn:VLVN}), 
the CKM matrix for leptons amounts to the diagonalization matrix 
${\cal V}$ for the light neutrino mass matrix ${\cal N}$. 
From Eqs. (\ref{eqn:calN}) and (\ref{eqn:GG}) we have 
\begin{equation}
     {\cal N} = \frac {1}{y_N} \Lambda ^{(2)}_l \Gamma _6^{-1} 
                      {\cal N}_0 \,\Gamma _6^{-1} \Lambda ^{(2)}_l, 
\end{equation}
with 
\begin{equation}
     {\cal N}_0 = \Gamma _6 {\cal V}^{-1}_l 
                      \Gamma _6^{-1} N_0^{-1} \Gamma _6^{-1} 
                         {\cal V}_l \Gamma _6. 
\end{equation}
By observing that ${\cal N}_{0ij} = O(1)$, 
we find that the diagonalization matrix ${\cal V}$ is essentially 
determined by the hierarchical pattern of 
\begin{equation}
    \Lambda ^{(2)}_l \Gamma ^{-1}_6 \simeq 
       y^{(l)}_M \,x^{\alpha _5 + \gamma _5} \times {\rm diag} 
            \left( x^{\xi}, \quad 1, \quad 
                   x^{ - \delta _l} \right). 
\end{equation}
Concretely, we have 
\begin{equation} 
    V^L_{\rm CKM} = {\cal V}  =  
      \left( 
      \begin{array}{ccc}
        1 - O(x^{2\xi})  &  O(x^{\xi})  
                              &  O(x^{\xi + \delta _l})  \\
        O(x^{\xi})    &  1 - O(x^{2\delta _l})  
                              &  O(x^{\delta _l})  \\
        O(x^{\xi + \delta _l})   &  O(x^{\delta _l})  
                              &  1 - O(x^{2\delta _l})  
      \end{array}
      \right). 
\label{eqn:LCKM}
\end{equation} 
It follows that the neutrino flavor 
mixing between $\nu _{\mu }$ and $\nu _{\tau}$ is expressed as 
\begin{equation}
   \sin \theta ^L_{23} \simeq x^{\delta _l} 
\end{equation}
up to RG effects. 
The neutrino masses are given by 
\begin{equation}
   m_{\nu} = \frac {(v_u \,y^{(l)}_M)^2}{M_S \,y_N} 
                    \,x^{2\alpha _5 + 2\gamma _5} \times 
                 \left( O(x^{2\xi}), \quad O(1), \quad 
                  O(x^{-2\delta _l}) \right), 
\end{equation}
which correspond to $\nu _e$, $\nu _{\mu}$ and $\nu _{\tau}$, 
respectively.

\section{The unification gauge group}

Regarding the unification gauge group $G$, 
up to now it has only been assumed that $G$ includes both $SU(2)_R$ 
and the standard model gauge group $G_{SM}$. 
Let us consider a larger unification group $G$ such that 
some of the matter fields in Eq.(\ref{eqn:PhiA}) reside in the same 
irreducible representations of $G$. 
In such a case we have several equalities among the $\Gamma _A$ with 
$A = 1,2,\cdots,8$.
Specifically, when $G$ contains $SU(4)_{PS} \times SU(2)_L 
\times SU(2)_R$,\cite{Pati} 
quarks and leptons are unified. 
Matter fields $\Phi _A$ in Eq.(\ref{eqn:PhiA}) fall into five 
categories , 
\begin{eqnarray*} 
     ({\bf 4, 2, 1})   &:& \ \ Q, L, \\
     ({\bf 6, 1, 1})   &:& \ \ g, g^c,  \\
     ({\bf 4^*, 1, 2}) &:& \ \ U^c, D^c, N^c, E^c, \\
     ({\bf 1, 2, 2})   &:& \ \ H_u, H_d, \\
     ({\bf 1, 1, 1})   &:& \ \ S, 
\end{eqnarray*} 
under $SU(4)_{PS} \times SU(2)_L \times SU(2)_R$. 
Consequently, the relations 
\begin{equation}
   \Gamma_1 = \Gamma_5, \qquad 
   \Gamma_2 = \Gamma_6, \qquad 
   \Gamma_3 = \Gamma_4 
\end{equation}
hold. 
Furthermore, from Eq.(\ref{eqn:m33}) we have 
$y^{(l)}_M = y^{(q)}_M = 1$. 
From Eq.(\ref{eqn:dell}) it follows that 
\begin{equation}
   \delta _l = \min (\alpha _1, \ \alpha _7 - \xi) \leq \alpha _1. 
\end{equation}
This leads us to the important relation 
\begin{equation}
    \sin \theta ^L_{23} \gsim \sin \theta ^Q_{12} = \lambda, 
\end{equation}
up to RG effects. 
As pointed out in Ref.\cite{LCKM}, 
since matter fields of the third generation have large Yukawa couplings, 
it is important to investigate renormalization effects of 
the flavor mixings $\sin \theta ^Q_{23}$ and $\sin \theta ^L_{23}$ 
from the scale $M_S$ to the electroweak scale. 
Concretely, the RG evolution from the scale $M_U(=\langle S_0 \rangle)$ 
to the scale $\langle N^c_0 \rangle$ introduces a sizable effect only 
to the (3, 3) element of $M^{(q)}$ and  $M^{(l)}$. 
As a consequence, we have the RG-enhanced $\sin \theta ^Q_{23}$ and 
$\sin \theta ^L_{23}$. 
Since effective Yukawa couplings of the first and the second 
generations are very small, 
renormalization effects of $\sin \theta _{12}$ are rather small 
compared with those of $\sin \theta _{23}$. 
Numerically, it can be shown that $\sin \theta _{23}$ is 
enhanced by about a factor of 2 when approaching the scale 
$\langle N^c_0 \rangle$ from $M_U$. 
Then we obtain 
\begin{equation}
   \sin \theta ^L_{23}(M_Z) \gsim 2\lambda \simeq 0.44. 
\end{equation}
This is consistent with atmospheric neutrino data.
\cite{Atmos} 
Furthermore, in the case of 
$G = SU(4)_{PS} \times SU(2)_L \times SU(2)_R$, 
the mass eigenvalues turn out to be 
\begin{equation}
   v_u \times \left( 
   O(x^{\alpha _1 + \gamma _1 +\alpha _2 + \gamma _2}), \qquad 
   O(x^{\gamma _1 + \gamma _2}), \qquad  
   O(1) \right) 
\label{eqn:massu}
\end{equation}
for up-type quarks, 
\begin{equation}
   v_d \, x^{\alpha _2 + \gamma _2} \times \left( 
   O(x^{\alpha _1 + \gamma _1}), \quad 
   O(x^{\gamma _1}), \quad 
   O(x^{- \delta _d}) \right) 
\end{equation}
for down-type quarks, 
\begin{equation}
   v_d \, x^{\alpha _1 + \gamma _1} \times \left( 
   O(x^{\alpha _2 + \gamma _2 + \xi}), \quad 
   O(x^{\gamma _2}), \quad 
   O(x^{- \delta _l}) \right) 
\end{equation}
for charged leptons and 
\begin{equation}
   m_{\nu} \simeq \frac {v_u^2}{M_S \,y_N} 
                 \, x^{2\alpha _1 + 2\gamma _1} \times 
                 \left( O(x^{2\xi}), \quad O(1), \quad 
                  O(x^{-2\delta _l}) \right) 
\label{eqn:massn}
\end{equation}
for neutrinos, where $\delta _d = \min (\alpha _2, \ \alpha _3)$ 
and $\delta _l = \min (\alpha _1, \ \alpha _7 - \xi)$.

If $G$ contains the traditional $SO(10)$, 
matter fields $\Phi _A$ are classified as 
\begin{eqnarray*} 
   ({\bf 16})  &:& \ \ Q, L, U^c, D^c, N^c, E^c, \\
   ({\bf 10})  &:& \ \ g, g^c, H_u, H_d, \\
   ({\bf  1})  &:& \ \ S, 
\end{eqnarray*} 
under $SO(10)$. 
It follows that 
\begin{equation}
   \Gamma_1 = \Gamma_2 = \Gamma_5 = \Gamma_6, \qquad 
   \Gamma_3 = \Gamma_4 = \Gamma_7. 
\end{equation}
In this case, however, it is impossible to break down the gauge group 
$SO(10)$ into $G_{SM}$ without introducing a higher representation 
field of $SO(10)$, 
which is not allowed in the level-one string theory. 
$SU(6) \times SU(2)_R$ is the largest unification gauge group 
from which we can implement the breakdown to $G_{SM}$ without 
introducing an additional field. 
When $G = SU(6) \times SU(2)_R$, 
matter fields $\Phi_A$ are classified as
\cite{QCKM}\cite{LCKM} 
\begin{eqnarray*} 
\Phi ({\bf 15, 1})  &:& \ \ Q, L, g, g^c, S, \\
\Psi ({\bf 6^*, 2}) &:& \ \ U^c, D^c, N^c, E^c, H_u, H_d. 
\end{eqnarray*} 
Thus we obtain 
\begin{equation}
   \Gamma_1 = \Gamma_3 = \Gamma_4 = \Gamma_5 = \Gamma_8, \qquad 
   \Gamma_2 = \Gamma_6 = \Gamma_7. 
\end{equation}
Then, in the above mass eigenvalues Eqs.(\ref{eqn:massu}) to 
(\ref{eqn:massn}) we have $\delta _d = \min (\alpha _1, \ \alpha _2)$ 
and $\delta _l = \min (\alpha _1, \ \alpha _2 - \xi)$. 
Since the traditional $SU(5)$ gauge group does not include 
$SU(2)_R$, 
the $SU(5)$ GUT is not in accord with the present framework.

\section{Summary}

The CKM matrix exhibits the disparity between the diagonalization 
matrices for up-type quarks (leptons) and down-type quarks 
(leptons) in $SU(2)_L$-doublets. 
With the assumption that the unification gauge 
group $G$ includes $SU(2)_R$, 
we explored the origin of the disparity purely within the 
effective field theory below the unification scale $M_U$. 
In the supersymmetric $E_6$-type models, 
quarks and leptons mix with extra particles 
beyond the MSSM. 
It should be noted that these extra particles contain triplet and 
doublet Higgs superfields with odd $R$-parity in each generation. 
We showed that the disparity occurs when these extra-particle mixings 
are large.

For down-type quarks there appear mixings between $g^c$ 
and $D^c$, which are $SU(2)_L$-singlets. 
This type of the mixing does not occur for up-type quarks. 
When the mixings between $g^c$ and $D^c$ are large, 
the disparity arises in the diagonalization matrices from the 
mixings. 
Namely, a non-trivial CKM matrix is derived. 
By introducing the flavor symmetry $U(1) \times {\bf Z}_2$ 
and by assigning appropriate flavor charges to matter fields, 
we obtain a realistic solution which reproduces both the quark 
spectra and the CKM matrix. 
On the other hand, for leptons the mixings occur between 
$L$ and $H_d$, both of which are $SU(2)_L$-doublets. 
Therefore, as a result of this mixing there appears no disparity 
between the diagonalization matrices for charged leptons and 
neutrinos. 
However, the non-trivial texture of the CKM matrix of leptons 
emerges as a consequence of the seesaw mechanism with 
the hierarchical Majorana mass matrix for right-handed neutrinos. 
Appropriate assignments of flavor charges lead to a 
phenomenologically viable solution which explains both 
the lepton spectra and the large $\sin \theta ^L_{23}$.

With the assumption 
$G \supset SU(4)_{PS} \times SU(2)_L \times SU(2)_R$, 
we obtain phenomenologically viable relations among the 
fermion spectra, $V^Q_{\rm CKM}$ and $V^L_{\rm CKM}$. 
This suggests that the unification gauge group $G$ contains 
at least $SU(4)_{PS} \times SU(2)_L \times SU(2)_R$. 
The study of these relations provides an important clue 
to the gauge symmtry, the matter content, and 
the flavor symmetry at the unification scale.

\vspace{1cm}

\section*{Acknowledgements}

The author would like to thank Dr. N. Haba for valuable 
dicussions. 
This work was supported in part by the Grant-in-Aid for 
Scientific Research, Ministry of Education, Science 
and Culture, Japan (No.10140209 and No.10640256).



\end{document}